\documentclass[a4paper,11pt]{article}
\usepackage{pos}
\usepackage{graphicx} % Required for inserting images

\title{Radiative corrections to one- and two-meson tau decays for precise new physics tests}

\author*[a]{Pablo Roig}

\affiliation[a]{Departamento de Física, Centro de Investigación y de Estudios Avanzados del Instituto
Politécnico Nacional, Apdo. Postal 14-740, 07000 Ciudad de México, México.}    

\emailAdd{pablo.roig@cinvestav.mx}

\abstract{We review the radiative corrections to the $\tau \to P (P) \nu_\tau [\gamma]$ decays and their implications for several SM tests: lepton universality, CKM unitarity and non-standard interactions.}

\FullConference{10th International Conference on Quarks and Nuclear Physics (QNP2024)\\
8-12 July, 2024\\
Barcelona, Spain\\}

\date{}
\begin{document}

\maketitle

\section{Introduction}
The tau is the only lepton massive enough to decay into mesons, thus offering a clean environment to learn about hadronization of QCD currents. This traditional view has recently been complemented by exploiting its potential for stringent new physics (NP) tests \cite{Pich:2013lsa}.

Along these lines, here we will focus on surveying the radiative corrections (RadCors) to one- and two-meson tau decays, which enable more precise NP tests. The situation before our last paper \cite{Escribano:2023seb} is summarized in table \ref{Tab:RadCors4TauDecs}, where the precision of the branching ratio (BR) measurement and the NP applications are mentioned for each channel. Short-distance electroweak RadCors were computed by Marciano $\&$ Sirlin \cite{Sirlin:1977sv, Marciano:1988vm}. Electromagnetic RadCors require the inclusion of diagrams with both virtual (loops) and real photons (initial- and final-state radiation). I will illustrate this with the one-meson case and just state our results for two mesons in the concluding remarks.

%\begin{center}
\begin{table}[h!]
\begin{tabular}{c c c c}
\hline
Meson state & BR precision ($\%$) \cite{ParticleDataGroup:2022pth} & RadCor & Application\\
\hline
$\pi^-$ & $0.5$ & \cite{Decker:1994ea,Arroyo-Urena:2021nil,Arroyo-Urena:2021dfe} & LFU, NSI\\
$K^-$ & $1.4$ & \cite{Decker:1994ea,Arroyo-Urena:2021nil,Arroyo-Urena:2021dfe} & $V_{us}$, LFU, NSI\\
$\pi^-\pi^0$ & $0.4$ & \cite{Cirigliano:2001er, Cirigliano:2002pv,Flores-Baez:2006yiq,Miranda:2020wdg} & $\rho^{(')},a_\mu$, NSI\\
$K^-K^0$ & $2.3$ & x & $\rho^{(')}$, NSI\\
$\bar{K}^0\pi^-$ & $1.7$ & \cite{Antonelli:2013usa, Flores-Baez:2013eba} & $K^*,V_{us}$, CPV, NSI\\
$K^-\pi^0$ & $3.5$ & \cite{Antonelli:2013usa, Flores-Baez:2013eba} & $K^*,V_{us}$, NSI\\
$K^-\eta$ & $5.2$ & x & $K^*$, NSI\\
$\pi^-\pi^-\pi^+$ & $0.5$ & x & $a_1$\\ 
$\pi^-\pi^0\pi^0$ & $1.1$ & x & $a_1$\\ 
\hline
\end{tabular}
\caption{Summary of RadCors for semileptonic tau decay channels, before our paper \cite{Escribano:2023seb}, which completed them for two-meson tau decays (only the structure-independent one was known for the $K\pi$ modes).}\label{Tab:RadCors4TauDecs}
\end{table}
%\end{center}

\section{RadCors to one-meson tau decays: Motivation}
Our main observable of interest is $R_{\tau/P}\equiv\frac{\Gamma(\tau\to P\nu_\tau[\gamma])}{\Gamma(P\to\mu\nu_\mu[\gamma])}$, which tests lepton flavor universality (LU) for $P=\pi,K$, verifying if $g_\tau=g_\mu$, stemming from the SM gauge symmetry. Anomalies related to LU violations have attracted much attention recently in $B$ decays \cite{Albrecht:2021tul}. While the quark flavors involved are different, low-energy tests of LU currently allow for the most precise analyses of this fundamental property \cite{Bryman:2021teu}.
Specifically, our main RadCor, $\delta R_{\tau/P}$ \cite{Decker:1994ea}, is defined through $R_{\tau/P}=\Big|\frac{g_\tau}{g_\mu}\Big|^2R_{\tau/P}^{(0)}(1+\delta R_{\tau/P})$, where the leading order (LO) SM result, $R_{\tau/P}^{(0)}$ depends only on $M_\tau, m_\mu, m_P$, which results in a negligible numerical uncertainty.

Important phenomenological and theoretical reasons motivated our reanalysis of $\delta R_{\tau/P}$ \cite{Arroyo-Urena:2021nil,Arroyo-Urena:2021dfe}. First, lepton universality in purely lepton decays is verified within less than one $\sigma$ \cite{HFLAV:2019otj,HFLAV:2022esi}. Also, ATLAS and CMS washed away a long-standing tension from LEP on the ratios between lepton $W$ decays into $\tau$ flavor, over $\mu$ and e flavors \cite{ATLAS:2020xea, CMS:2022mhs}. While these two tests at the mil and six per mil level, respectively, reinforce LU at high precision; $R_{\tau/P}$ was seemingly deviating from LU with a significance in the range $[1.5,2]\sigma$, which deserved a close scrutiny. On the formal side, a cutoff was employed in ref.~\cite{Decker:1994ea} to regulate the loop integrals, axiomatic QFT properties were only fulfilled at LO, and the quoted uncertainties were as small as expected in $P\to\mu\nu_\mu$, but not in tau decays, where the dynamics is much more complicated by the prominent r\^ole of resonances.
As byproducts of our project, we will also quote the RadCors on the one-meson tau decays, and use them to perform CKM unitarity and non-standard interactions (NSI) tests. 
\section{RadCors to $P\to\mu\nu_\mu[\gamma]$ decays}
The structure-dependent (SD) part of these RCs is calculated unambiguously within Chiral Perturbation Theory \cite{Weinberg:1978kz,Gasser:1983yg,Gasser:1984gg}. The model-independent one was obtained by Kinoshita \cite{Kinoshita:1959ha} and the short-distance electroweak universal corrections is included in the $S_{EW}$ factor \cite{Sirlin:1977sv, Marciano:1988vm}. The SD piece was computed by Cirigliano and Rosell \cite{Cirigliano:2007xi,Cirigliano:2007ga} at two loops, and only two counterterms remained undetermined. A large number of colors ($N_C$) expansion of QCD in the resonance region was applied to ensure a short-distance behaviour in accord with the parton picture \cite{Ecker:1988te,Ecker:1989yg,Cirigliano:2006hb}, reducing thereby the associated uncertainty.
\section{RadCors to $\tau\to P\nu_\tau[\gamma]$ decays}
The same effective action for QCD in the large-$N_C$ limit is used here to evaluate the SD part, wherein we studied the real radiation in refs.~\cite{Guo:2010dv,Guevara:2013wwa,Guevara:2021tpy} and the virtual photon piece in \cite{Arroyo-Urena:2021nil,Arroyo-Urena:2021dfe}. In analogy to the real photon case, both the vector and axial-vector currents contribute and, under chiral symmetry, only three form factors appear, one being directly related to the electromagnetic $P$ meson form factor (known to extremely good accuracy). The remaining two have very simple expressions when chiral symmetry breaking corrections are neglected and we stick to the lowest-lying LO contribution in the large $N_C$ expansion. Namely,
\begin{equation}\label{eq:FFs}
F_V^P(W^2,k^2)=\frac{-N_CM_V^4}{24\pi^2F_P(k^2-M_V^2)(W^2-M_V^2)}\,,\quad 
F_A^P(W^2,k^2)=\frac{F_P}{2}\frac{M_A^2-2M_V^2-k^2}{(k^2-M_V^2)(W^2-M_A^2)}\,,
\end{equation}
where $W^2$ and $k^2$ are the $W$ and $\gamma$ (which closes the loop, with the $\tau$ and $P$) virtualities, respectively, and the meson masses correspond to their value in the aforementioned limit, in which also the $P$ decay constant, $F_P$, is defined. Our setting warrants good ultraviolet behaviour of two- and three-point Green functions \cite{Cirigliano:2006hb,Kampf:2011ty,Roig:2013baa}. We will estimate our model-dependent uncertainty by also computing our RadCors with short-distance constraints corresponding only to two-point Green functions in accordance with QCD asymptotics \cite{Ecker:1988te,Ecker:1989yg}. We will find that this error is subdominant, suggesting that more refined form factors are not needed presently.
\section{Calculation of $R_{\tau/P}$}
As a first check, we confirm the results by Decker and Finkemeier \cite{Decker:1994ea} for the structure-independent (SI) part of the RadCor, $\delta R_{\tau/P}|_{SI}$, which is $1.05\%$ (uncertainties not quoted are always smaller than the accuracy shown) for $P=\pi$ and $1.67\%$ for $P=K$. Real (r) radiation is negligible in $P$ decays \cite{Cirigliano:2007xi,Cirigliano:2007ga}, but not for the $\tau$, where it amounts to $0.15\%$ ($P=\pi$) and $(0.18\pm0.15)\%$ ($P=K$).
The virtual (v) SD RadCor reads $\delta_{\pi\mu}|_{vSD}=(0.54\pm0.12)\%$ and $\delta_{K\mu}|_{vSD}=(0.43\pm0.12)\%$\cite{Cirigliano:2007xi,Cirigliano:2007ga} in $P$ decays and $\delta_{\tau\pi}|_{vSD}=(-0.48\pm0.56)\%$, $\delta_{\tau K}|_{vSD}=(-0.45\pm0.57)\%$ \cite{Arroyo-Urena:2021nil,Arroyo-Urena:2021dfe} for $\tau$, yielding \cite{Arroyo-Urena:2021nil,Arroyo-Urena:2021dfe} $\delta R_{\tau/\pi}|_{vSD}=(-1.02\pm0.57)\%$, $\delta R_{\tau/K}|_{vSD}=(-0.88\pm0.58)\%$. The different sign of the vSD corrections to numerator and denominator of $R_{\tau/K}$ make their combined effect comparable in size (although with opposite sign) to the SI part. The partial cancellation among these two contributions will make that the final result (particularly for $\pi$) is quite close to the rSD Rad Cor, which is the smallest among the three components. The uncertainty of the overall correction will be saturated by that of the vSD piece (see table \ref{Tab:Results}), which basically comes from the one associated to the $\mu$-dependence of the loop integrals.
\section{Results}
Our results for the RadCor to $R_{\tau/P}$ are summarized in table \ref{Tab:Results}.
\begin{table}[h!]
\begin{tabular}{c c c}
\hline
Contribution & $\delta R_{\tau/\pi}$ & $\delta R_{\tau/K}$\\
\hline
SI & $+1.05\%$ & $+1.67\%$ \\
rSD & $+0.15\%$ & $(+0.18\pm0.15)\%$ \\
vSD & $(-1.02\pm0.57)\%$ & $(-0.88\pm0.58)\%$\\
\hline
Total & $(+0.18\pm0.57)\%$ & $(+0.97\pm0.58)\%$\\
\hline
\end{tabular}
\caption{Our results for the RadCor $\delta R_{\tau/P}$ \cite{Arroyo-Urena:2021nil,Arroyo-Urena:2021dfe}.}\label{Tab:Results}
\end{table}
We point out that our agreement for the central values with \cite{Decker:1994ea} is a mere coincidence. Particularly, detailed comparisons are difficult since their cutting-off the integrals splits unphysically long- and short-distance regimes and leaves an artificial dependence on the associated scale in the results.
\section{Applications}
\subsection{RadCors to $\Gamma(\tau\to P\nu_\tau[\gamma])$}
As a first application, we give the RadCor (SI+SD) entering the one-meson tau decays
\begin{equation}
\Gamma(\tau\to P\nu_\tau[\gamma]))\frac{G_F^2|V_{uD}|^2F_P^2}{8\pi}M_\tau^3\left(1-\frac{m_P^2}{M_\tau^2}\right)S_{EW}(1+\delta_{\tau P})\,.
\end{equation}
with $\delta_{\tau \pi}=(-0.24\pm0.56)\%$ and $\delta_{\tau \pi}=(-0.15\pm0.57)\%$
\subsection{LU test}
As our main result, we get
\begin{equation}
    R_{\tau/P}\equiv \frac{\Gamma(\tau\to P\nu_\tau[\gamma])}{\Gamma(P\to\mu\nu_\mu[\gamma])}\Bigg|\frac{g_\tau}{g_\mu}\Bigg|_P^2 R_{\tau/P}^0(1+\delta R_{\tau/P})\,.
\end{equation}
Using our RadCors and PDG input, we get
\begin{equation}
\Big|\frac{g_\tau}{g_\mu}\Big|_\pi=0.9964\pm0.0038\,,\quad \Big|\frac{g_\tau}{g_\mu}\Big|_K=0.9857\pm0.0078\,,
\end{equation}
improving agreement on this LU test with respect to \cite{HFLAV:2019otj}. Agreement within one standard deviation is reached for the $\pi$. The difference in the Kaon case is only mildly diminished (to $1.8\sigma$) because uncertainties are mostly statistical.
\subsection{CKM unitarity test in $\Gamma(\tau\to K\nu_\tau[\gamma])/\Gamma(\tau\to \pi\nu_\tau[\gamma])$}
The first Cabibbo unitarity test uses the ratio among the one meson tau decays
\begin{equation} \label{eq_K/pi}
\frac{\Gamma(\tau\to K\nu_\tau[\gamma])}{\Gamma(\tau\to \pi\nu_\tau[\gamma])}=\Bigg|\frac{V_{us}}{V_{ud}}\Bigg|^2\frac{F_K^2}{F_\pi^2}\frac{\left(1-m_K^2/M_\tau^2\right)^2}{\left(1-m_\pi^2/M_\tau^2\right)^2}(1+\delta)\,,    
\end{equation}
with $\delta=(+0.10\pm0.80)\%$~\footnote{Although this is our published value~\cite{Arroyo-Urena:2021nil,Arroyo-Urena:2021dfe}, it is extremely conservative, as it neglects the large correlation on the renormalization scale dependence of numerator and denominator in eq.~(\ref{eq_K/pi}). We thank Swagato Banerjee and Alberto Lusiani for insisting on the convenience that we provided the uncertainty accounting for this correlation. In that case, the main uncertainty comes from the difference of the results obtained with the original resonance Lagrangian of ref.~\cite{Ecker:1988te} or its extended version \cite{Cirigliano:2006hb,Kampf:2011ty}. In this way, the uncertainty on $\delta$ drops from $0.80\%$ to $(0.22\oplus0.24)\%=0.33\%$, where we have neglected the (reasonably small this time) correlation between numerator and denominator.}. This gives \cite{ParticleDataGroup:2022pth,FlavourLatticeAveragingGroupFLAG:2021npn} $\Big|\frac{V_{us}}{V_{ud}}\Big|=0.2288\pm0.0020$, that is $2.1\sigma$ away from CKM unitarity, considering $V_{ud}= 0.97373\pm0.000031$ \cite{Hardy:2020qwl}. This result is much less precise than $\Big|\frac{V_{us}}{V_{ud}}\Big|=0.2291\pm0.0009$ \cite{Seng:2021nar}, obtained with kaon semileptonic decays. Again we are limited by the statistical uncertainties of tau decays measurements.
\subsection{CKM unitarity test in $\Gamma(\tau\to K\nu_\tau[\gamma])$}
A similar test can be carried out using just the decays into kaons, via
\begin{equation}
\Gamma(\tau\to K\nu_\tau[\gamma])=\frac{G_F^2F_K^2}{8\pi}|V_{us}|^2M_\tau^3\left(1-\frac{m_K^2}{M_\tau^2}\right)^2 S_{EW}(1+\delta_{\tau K})\,,
\end{equation}
yielding $|V_{us}|=0.2220\pm0.0018$, $2.6\sigma$ away from unitarity using the  same $V_{ud}$ quoted previously \cite{Hardy:2020qwl}. The similar test with Kaon semileptonic decays gives $|V_{us}|=0.2231\pm0.0006$, three times more precise thanks to statistics.
\subsection{NSI constraints from $\Gamma(\tau\to P\nu_\tau[\gamma])$}
Finally, we can constrain NSI \cite{Cirigliano:2009wk} from the one-meson tau decays, where their effect implies the correction $\delta_{\tau P}\to \delta_{\tau P}+2\Delta_{\tau P}$, with
\begin{equation}
\Delta_{\tau P}=\epsilon_L^\tau-\epsilon_L^e-\epsilon_R^\tau-\epsilon_R^e-\frac{m_P^2}{M_\tau(m_u+m_D)}\epsilon^\tau_P.
\end{equation}
The limits $\Delta_{\tau \pi}=(-0.15\pm0.72)\times10^{-2}$ and $\Delta_{\tau K}=(-0.36\pm1.18)\times10^{-2}$~\footnote{These are obtained in the $\overline{MS}$-scheme at $\mu=2$ GeV. Similar results were achieved in refs. \cite{Cirigliano:2018dyk,Gonzalez-Solis:2020jlh,Cirigliano:2021yto}.} correspond to NP scales beyond $\sim 3$TeV, for fundamental couplings of SM strength.
\section{Conclusions (also on di-meson tau decays and NP tests through them)}
Our improved RadCors in Table \ref{Tab:Results} enable reliable tests of LU, agreeing with that essential property within one sigma, for the pion case. For Kaons, the difference is still $1.8\sigma$, and the impact of our improvement is modest, since the result is statistically dominated. We have also performed Cabibbo unitarity tests (using the ratio amond both one-meson tau decays, or only the Kaon mode) which depart between $[2.1,2.6]\sigma$ from unitarity. Our uncertainty is at least thrice the one of the tests using Kaon semileptonic decays, which calls for improved measurements of meson tau decays at Belle-II and future facilities. Our constraints on NSI are less powerful than those from Kaons, but still help constraining flat directions in the NP parameter space. Our results have been incorporated in the HFLAV'22 review \cite{HFLAV:2022esi}.

We finish summarizing the results for two mesons tau decays. Using dispersive form factors for the non-radiative decays \cite{GomezDumm:2013sib,Gonzalez-Solis:2019iod,Jamin:2001zq,Escribano:2013bca,Escribano:2014joa,Escribano:2016ntp}, our analysis \cite{Escribano:2023seb} (see also \cite{Guevara:2016trs,GutierrezSantiago:2020bhy}) determines the following RadCors (in $\%$)
\begin{eqnarray}
\delta^{K^-\pi^0}=-0.009^{+0.010}_{-0.118}\,,\delta^{\bar{K}^0\pi^-}=-0.166^{+0.100}_{-0.157}\,,\delta^{K^-K^0}=-0.030^{+0.032}_{-0.180}\,,\nonumber\\
\delta^{\pi^-\pi^0}=-0.186^{+0.114}_{-0.203}\,,\delta^{K^-\eta}=-0.026^{+0.029}_{-0.163}\,,\delta^{K^-\eta^\prime}=-0.304^{+0.422}_{-0.185}\,,
\end{eqnarray}
which -for the $K\pi$ case- halve previous uncertainties. The corresponding implications for NSI were updated (see also e. g. refs.~\cite{Garces:2017jpz,Cirigliano:2017tqn,Miranda:2018cpf,Rendon:2019awg,Gonzalez-Solis:2019lze,Miranda:2020wdg,Masjuan:2023qsp,Castro:2024prg}) in our paper \cite{Escribano:2023seb}.
\acknowledgments
It is my pleasure to thank the QNP24 organizing committees. I am indebted to Rafel Escribano, for pleasant collaborations over the years enriched with a deep human component, and for financial support through the Ministerio de Ciencia e Innovación under Grant No. PID2020–112965 GB-I00, and by the Departament de Recerca i Universitats from Generalitat de Catalunya to the Grup de Recerca ``Grup de Física Teòrica UAB/IFAE'' (Codi: 2021 SGR 00649). I acknowledge partial Conahcyt support, specifically from project CB2023-2024-3226. I thank wholeheartedly the hospitality of the SM group at IFAE -where this contribution was written- which I was extremely glad to visit again, eleven years after I left at the end of my first postdoc.

\end{document}